\documentclass[12pt]{article}
\usepackage{amsfonts}
\usepackage{amsmath, amsthm}
\usepackage{epsfig}
\usepackage{algorithm}
\usepackage{algorithmic}
\usepackage{epic}
\usepackage[a4paper]{geometry}
\usepackage{enumitem}
\usepackage{amsmath,verbatim,color,amssymb,epsfig}
\usepackage{bm}
\usepackage[flushleft]{threeparttable}
\usepackage{amsfonts}
\usepackage{epsfig}
\usepackage{changepage}
\usepackage{multirow}
\usepackage{graphicx,booktabs}
\usepackage{array}
\usepackage[table]{xcolor}
\definecolor{Gray}{gray}{0.80}
\usepackage{lscape}
\usepackage[normalem]{ulem}
\usepackage[colorlinks=true,urlcolor=blue,citecolor=purple,linkcolor=blue,bookmarks=true]{hyperref}
\usepackage{caption,subcaption}
\begin{document}
\def\eqx"#1"{{\label{#1}}}
\def\eqn"#1"{{\ref{#1}}}

\makeatletter 
\@addtoreset{equation}{section}
\makeatother  

\def\yincomment#1{\vskip 2mm\boxit{\vskip 2mm{\color{red}\bf#1} {\color{blue}\bf --Yin\vskip 2mm}}\vskip 2mm}
\def\lincomment#1{\vskip 2mm\boxit{\vskip 2mm{\color{blue}\bf#1} {\color{black}\bf --Lin\vskip 2mm}}\vskip 2mm}
\def\squarebox#1{\hbox to #1{\hfill\vbox to #1{\vfill}}}
\def\boxit#1{\vbox{\hrule\hbox{\vrule\kern6pt
          \vbox{\kern6pt#1\kern6pt}\kern6pt\vrule}\hrule}}

\def\theequation{\thesection.\arabic{equation}}
\newcommand{\ds}{\displaystyle}

\newcommand{\bJ}{\mbox{\bf J}}
\newcommand{\bF}{\mbox{\bf F}}
\newcommand{\bM}{\mbox{\bf M}}
\newcommand{\bR}{\mbox{\bf R}}
\newcommand{\bZ}{\mbox{\bf Z}}
\newcommand{\bX}{\mbox{\bf X}}
\newcommand{\bx}{\mbox{\bf x}}
\newcommand{\bww}{\mbox{\bf w}}
\newcommand{\bQ}{\mbox{\bf Q}}
\newcommand{\bH}{\mbox{\bf H}}
\newcommand{\bh}{\mbox{\bf h}}
\newcommand{\bz}{\mbox{\bf z}}
\newcommand{\br}{\mbox{\bf r}}
\newcommand{\ba}{\mbox{\bf a}}
\newcommand{\be}{\mbox{\bf e}}
\newcommand{\bG}{\mbox{\bf G}}
\newcommand{\bB}{\mbox{\bf B}}
\newcommand{\bb}{\mbox{\bf b}}
\newcommand{\bA}{\mbox{\bf A}}
\newcommand{\bC}{\mbox{\bf C}}
\newcommand{\bI}{\mbox{\bf I}}
\newcommand{\bD}{\mbox{\bf D}}
\newcommand{\bU}{\mbox{\bf U}}
\newcommand{\bc}{\mbox{\bf c}}
\newcommand{\bd}{\mbox{\bf d}}
\newcommand{\bs}{\mbox{\bf s}}
\newcommand{\bS}{\mbox{\bf S}}
\newcommand{\bV}{\mbox{\bf V}}
\newcommand{\bv}{\mbox{\bf v}}
\newcommand{\bW}{\mbox{\bf W}}
\newcommand{\bY}{\mathbf{ Y}}
\newcommand{\bw}{\mbox{\bf w}}
\newcommand{\bg}{\mbox{\bf g}}
\newcommand{\bu}{\mbox{\bf u}}
\newcommand{\mI}{\mbox{I}}

\newcommand{\bch}{\color{blue}\it}
\newcommand{\ech}{\color{black}\rm}

\def\bb{{\bf b}}

\newcommand{\bcU}{\boldsymbol{\cal U}}
\newcommand{\bbeta}{\boldsymbol{\beta}}
\newcommand{\bdelta}{\boldsymbol{\delta}}
\newcommand{\bDelta}{\boldsymbol{\Delta}}
\newcommand{\boldeta}{\boldsymbol{\eta}}
\newcommand{\bxi}{\boldsymbol{\xi}}
\newcommand{\bGamma}{\boldsymbol{\Gamma}}
\newcommand{\bSigma}{\boldsymbol{\Sigma}}
\newcommand{\balpha}{\boldsymbol{\alpha}}
\newcommand{\bOmega}{\boldsymbol{\Omega}}
\newcommand{\btheta}{\boldsymbol{\theta}}
\newcommand{\bepsilon}{\boldsymbol{\epsilon}}
\newcommand{\bmu}{\boldsymbol{\mu}}
\newcommand{\bnu}{\boldsymbol{\nu}}
\newcommand{\bgamma}{\boldsymbol{\gamma}}
\newcommand{\btau}{\boldsymbol{\tau}}
\newcommand{\bTheta}{\boldsymbol{\Theta}}

\newtheorem{thm}{Theorem}
\newtheorem{lem}{Lemma}[section]
\newtheorem{rem}{Remark}[section]
\newtheorem{cor}{Corollary}[section]
\newcolumntype{L}[1]{>{\raggedright\let\newline\\\arraybackslash\hspace{0pt}}m{#1}}
\newcolumntype{C}[1]{>{\centering\let\newline\\\arraybackslash\hspace{0pt}}m{#1}}
\newcolumntype{R}[1]{>{\raggedleft\let\newline\\\arraybackslash\hspace{0pt}}m{#1}}

\newcommand{\tabincell}[2]{\begin{tabular}{@{}#1@{}}#2\end{tabular}}

\baselineskip=24pt
\begin{center}
{\Large \bf BOP2-DC: Bayesian optimal phase II designs with dual-criterion decision making}
\end{center}

\vspace{2mm}
\begin{center}
{Yujie Zhao$^1$, Daniel Li$^2$, Rong Liu$^2$, and Ying Yuan$^1$}
\end{center}

\begin{center}

$^1$ Department of Biostatistics, The University of Texas MD Anderson Cancer Center\\
Houston, Texas 77030, USA \\
$^2$Bristol Myers Squibb, Berkeley Heights, New Jersey 08901, USA \\
\vspace{2mm}

\end{center}
\noindent{\bf Abstract}

The conventional phase II trial design paradigm is to make the go/no-go decision based on the hypothesis testing framework. Statistical significance itself alone, however, may not be sufficient to establish that the drug is clinically effective enough to warrant confirmatory phase III trials. We propose the Bayesian optimal phase II trial design with dual-criterion decision making (BOP2-DC), which incorporates both statistical significance and clinical relevance into decision making. Based on the posterior probability that the treatment effect reaches the lower reference value (statistical significance) and the clinically meaningful value (clinical significance), BOP2-DC allows for go/consider/no-go decisions, rather than a binary go/no-go decision, and it is optimized to maximize the probability of a go decision when the treatment is effective or minimize the sample size when the treatment is futile. BOP2-DC is highly flexible and accommodates various types of endpoints, including binary, continuous, time-to-event, multiple, and co-primary endpoints, in single-arm and randomized trials. Simulation studies show that the BOP2-DC design yields desirable operating characteristics. The software to implement BOP2-DC is freely available at \url{www.trialdesign.org}.
\vspace{0.5cm}

\noindent{KEY WORDS:}  Bayesian adaptive design, phase II trials, go/consider/no-go decision, optimal design.

\section{Introduction}
The primary objective of phase II or proof-of-concept trials is to evaluate the efficacy of the investigational drug and make a go/no-go decision on whether the drug warrants further development. The conventional paradigm is to make the go/no-go decision based on the hypothesis testing framework with the null hypothesis that the drug is ineffective compared to a historical or concurrent control. That is, making a go decision if the efficacy reaches statistical significance; otherwise making a no-go decision. An example is Simon’s optimal two stage design (1989), which minimizes the expected or maximum sample size under the null hypothesis, while controlling the type I and type II error rates. 

Statistical significance alone, however, may not be sufficient to establish that a drug is sufficiently clinically effective to warrant confirmatory phase III trials.  It is well known that a statistically significant efficacy improvement may be too small to be considered clinically significant, for example, when the sample size is large or the inter-subject variability is small. Conversely, a clinically significant improvement may not reach a statistical significance, for example, when the study is underpowered. To address this issue, the dual-criterion decision-making framework was developed to take both statistical significance and clinical relevance into consideration. Sargent et al. (2001) introduced the additional uncertainty region in decision-making in phase II trials. Fisch et al. (2014) proposed a Bayesian proof-of-concept design that evaluates both evidence of significance and evidence of relevance in efficacy and discussed the three possible outcomes (i.e., go/indeterminate/stop) in decision-making. Frewer et al. (2016) proposed a frequentist decision-making framework by comparing the boundaries of confidence interval of estimators for treatment effect with two-level target product profiles. Kirby et al. (2017) compared five decision-making approaches for clinical efficacy including hypothesis testing and single and dual-criterion decision-making. Bertsche et al. (2017) optimized proof-of-concept design based on the use of historical information for time-to-event endpoints. Dunyak et al. (2017) integrated the dose estimation into a dual-criterion decision-making framework. Pulkstenis et al. (2017) and Quan et al. (2020) discussed the impact of sample size in different choices of target product profiles and decision-making thresholds for binary and continuous endpoints.

We develop the Bayesian optimal phase II trial design with dual-criterion decision making (BOP2-DC) with several new contributions. First, BOP2-DC is highly flexible and can handle various types of endpoints, including binary, continuous and time-to-event endpoints, as well as multiple and co-primary endpoints. In contrast, most existing due-criterion designs focus on a single binary endpoint. Second, BOP2-DC provides a rigorous and coherent approach to optimize design parameters to ensure that the design has desirable operating characteristics (e.g., control the false go rate and optimize the correct go rate), whereas most existing publications provide little guidance on choosing design parameters. Third, BOP2-DC accommodates both single-arm and randomized trials. Lastly, we develop a web app with a graphical user interface available on \url{www.trialdesign.org} that allows investigators to apply BOP2-DC in practice, which includes generating decision tables, operating characteristics of the design, and a protocol template.

BOP2-DC is an extension of the Bayesian optimal phase II (BOP2) design (Zhou et al., 2017), a flexible phase II design that handles various types of endpoints. 
BOP2 takes a binary-decision framework and recommends go/no-go decision, with software freely available on www.trialdesign.org. BOP2 has been increasingly in practice. At MD Anderson Cancer Center, over 40 ongoing trials use BOP2, and to our knowledge, many pharmaceutical companies employed BOP2. One most common query we have received from practitioners, particularly from industry biostatisticians, is that whether BOP2 can be extended to the dual-criterion decision-making framework to take both statistical significance and clinical relevance into consideration. By allowing ``consider" decision, investigators can make the final go/no-go decision (i.e., phase III trials) based on  the totality of the data (efficacy, safety, and PK and PD biomarkers).  Such flexibility is highly desirable in practice because the decision of whether proceed to phase III trial often depends on various considerations, not merely the primary endpoint. As shown below, the extension of BOP2 to BOP2-DC is not trivial. It requires new statistical considerations on decision rules and design parameter optimization. 


The remainder of this article is organized as follows. In Section 2, we describe our probability models and the decision-making algorithm for single-arm and randomized studies.  In Section 3, we evaluate the operating characteristics of BOP2-DC using simulations. We conclude with a brief discussion in Section 4.\\

\section{Single-arm study}

\subsection{Probability model}
Consider a phase II trial with $Y$ as the primary endpoint, which can be a binary, continuous, or time-to-event endpoint. In trials involving multiple endpoints or co-primary endpoints, $Y$ is a vector including multiple primary endpoints, as described later. Let $n$ denote the interim sample size and $\theta$ generically denote the treatment effect, which could be a scalar or vector. In what follows, we describe the statistical model and posterior inference of $\theta$ for each type of endpoint.

\subsubsection{Binary endpoint}    
A typical example of binary endpoints for phase II trials is tumor response rate, with $\theta$ representing the objective response rate (ORR).  Let $Y$ denote the number of patients with response among $n$ patients. We assume a beta-binomial model 
\begin{eqnarray*}
Y  &\sim& Binomial (\theta) \\
\theta &\sim& Beta(a_1, b_1),
\end{eqnarray*}
where $a_1$ and $b_1$ are hyperparameters. Typically, we set $a_1$ and $b_1$ small values (e.g., $a_1=b_1=0.1$) so that the resulting prior is vague with a prior effective sample size of $a_1+b_1$. The posterior distribution of $\theta$ is given by
$$\theta|X \sim Beta(a_1+Y, b_1+n-Y).$$

\subsubsection{Continuous endpoint}
Examples of continuous (or approximately continuous) endpoints include pharmacokinetics/pharmacodynamics outcomes, expression of biomarkers, and quality of life. Let $Y_i$ denote the endpoint for the $i$th patient. We assume that 
$$Y_1,  ..., Y_n \stackrel{\text{i.i.d}}{\sim} N(\theta,\sigma^2),$$
where $\theta$ is the mean treatment effect, and $\sigma^2$ is the variance. We assign $(\theta,\sigma^2)$ a normal-inverse-gamma prior 
 $$\theta | \sigma^2 \sim N(\theta_0,\frac{\sigma^2}{n_0}), \quad {\sigma^2} \sim IG(a_2, b_2), $$ 
where $IG(\cdot)$ denotes an inverse gamma distribution, and $\theta_0$, $n_0$, $a_2$ and $b_2$ are hyperparameters. 
The posterior distribution for $(\theta, \sigma^2)$ is
\begin{equation*}
\begin{aligned}
\theta|\sigma^2,Y_1,  ..., Y_n &\sim N(\frac{n\bar{Y}+n_0\theta_0}{n+n_0},\frac{\sigma^2}{n+n_0})\\
{\sigma^2} |Y_1,  ..., Y_n &\sim IG(a_2+\frac{n}{2},b_2+\frac{1}{2}\sum_{i=1}^n (Y_i-\bar{Y})^2+\frac{1}{2(\frac{1}{n}+\frac{1}{n_0})}(\bar{Y}-\theta_0)^2),  \\
\end{aligned}
\end{equation*}
where $\bar{Y} = \sum_{i=1}^n Y_i$. In most applications, it is desirable to use a vague prior to limit the influence of the prior on the posterior of $\theta$ by setting $n_0, a$ and $b$ at small values, e.g., $n_0=10^{-3}$ and $a=b=10^{-6}$.  For $\theta_0$, we can set it at the best prior estimate of $\theta$. Accurate specification of $\theta_0$ is not critical when $n_0$ is small, as $n_0$ represents the prior effective sample size. When there is reliable prior information, we can set $n_0$ at an appropriate value to incorporate the prior information. For example, setting $n_0=10$ means that the prior information is equivalent to 10 patients.

\subsubsection{Time-to-event endpoint}
Commonly used time-to-event endpoints include progression-free survival (PFS) and overall survival (OS). Let $Y_i$ denote the event time, $U_i$ is the censoring time, and define the observed time $T_i=min(Y_i,U_i)$ and the censoring indicator $\delta_i = I(Y_i > U_i)$,  for the $i$th patient. Let $\theta$ denote the median PFS or OS, and define $\tilde{\theta} = \theta/ ln(2)$. We assume the following exponential-inverse-gamma model:
\begin{eqnarray*}
Y_i  &\sim& Exp(\tilde{\theta}) \\
\tilde{\theta} &\sim& IG(a_3, b_3).
\end{eqnarray*}
We set hyperparameters $a_3$ and $b_3$ at small values (e.g., $a_3=b_3=10^{-6}$) to obtain a vague prior.  The posterior distribution of $\tilde{\theta}$ is given by
$$\tilde{\theta}|Y_1, ...,Y_n \sim IG(\gamma+n-\sum_{i=1}^n \delta_i,\beta+\sum_{i=1}^N {T_i}).$$ 
Accordingly, the posterior of $\theta$ is 
$$\theta|Y_1, ..., Y_n \sim IG(\gamma+n - \sum_{i=1}^n \delta_i, (\beta+\sum_{i=1}^N {T_i})ln(2)).$$ 
The assumption that $Y_i$ follows an exponential distribution is strong, however, numerical studies show that, for the purpose of futility monitoring, the performance of this probability model is remarkably robust to the violation of this assumption (Thall et al. 2005 , Zhou et al. 2020).

\subsubsection{Multiple and co-primary endpoints}
Following the terminology of the US Food and Drug Administration (FDA) guidance (2017), we use multiple endpoints to denote the case that demonstration of a treatment effect on at least one of several primary endpoints is sufficient to establish clinical benefit, and co-primary endpoints to denote the case that demonstration of treatment effects on all endpoints is necessary to establish clinical benefit. An example of multiple points is a phase II endometrial cancer trial reported by Moore et al. (2015), where $Y_1$ is the ORR and $Y_2$ is an indicator of 6-month event-free survival (EFS6). The treatment is regarded clinically beneficial if either of the endpoints is reached. An example of co-primary endpoints is a trial jointly monitoring safety and efficacy, where $Y_1$ is a binary safety endpoint, and $Y_2$ is a binary efficacy endpoint. Establishing clinical benefit of the treatment requires that both safety and efficacy endpoints are reached. 
For ease of exposition, we focus on the case that multiple endpoints or co-primary endpoints $Y$ consist of two categorical endpoints, i.e., $Y = (Y_1, Y_2)$, where $Y_1$ and $Y_2$ are categorical variables with $R$ and $Q$ categories, respectively. Extension to more than two categorical endpoints is straightforward. 

The joint distribution of $(Y_1, Y_2)$ can be represented by a categorical variable $Z$ with $K = R \times Q$ categories. For example, when $Y_1$ and $Y_2$ are binary response and toxicity endpoints, respectively, $Z$ has $K=4$ categories: (response, toxicity), (response, no toxicity), (no response, toxicity), and (no response, no toxicity). Let $X_k$ denote the number of patients with $Z$ being in the $k$th category, and define $\bm{X} = (X_1, \cdots, X_K)$. We model $Z$ using a multinomial-Dirichlet model:
\begin{eqnarray*}
\bm{X}  &\sim&  Multinomial(p_{1}, ..., p_{K}) \\
p_{1},p_{2}, ...,p_{K} &\sim& Dirichlet(\alpha_{1}, ...,\alpha_{K}),
\end{eqnarray*}
where $p_k$ denotes the probability of $Z$ being in the $k$th category, and $\alpha_{1}, ...,\alpha_{K}$ are hyperparameters.  The posterior of $\bm{p}=(p_{1},p_{2}, ...,p_{K}) $ is 
$$\bm{p} | \bm{X} \sim Dirichlet(\alpha_{1}+X_{1},\alpha_{2}+X_{2},...,\alpha_{K}+X_{K}). $$ 
As $\sum_{k=1}^K \alpha_k$ represents the prior effective sample size, we set $\alpha_k$'s at small values, e.g., with $\sum_{k=1}^K \alpha_k=1$, to obtain a vague prior.

For multiple and co-primary endpoints, the parameter of clinical interest typically is not $\bm{p}$, but a linear combination of $\bm{p}$. For example, when the multiple endpoints are ORR and EFS6, $p_{1}, ..., p_{4}$ are the probabilities of a patient belonging to four possible categories: (response, EFS6), (response, not EFS6), (no response, EFS6), and  (no response, not EFS6). The efficacy parameter of clinical interest is $\theta = (\theta_1, \theta_2)$, where $\theta_1= p_1 + p_2$ is ORR, and $\theta_2=p_1 + p_3$ is EFS6. In the case with toxicity and response as co-primary endpoints, $p_{1}, ..., p_{4}$ are the probabilities of a patient belonging to four possible categories: (response, toxicity), (response, no toxicity), (no response, toxicity), and  (no response, no toxicity). The efficacy parameter of clinical interest is $\theta = (\theta_1, \theta_2)$, where $\theta_1= p_1 + p_2$ is ORR, and $\theta_2=p_1 + p_3$ is the toxicity rate. The inference of $\theta$ is facilitated by the following distribution property (Zhou et al., 2017). Let $\theta_j = \bm{b_jp^T}$ denote a linear combination of $\bm{p}$, where $\bm{b_j}$ is a $1\times K$ vector with elements of 0 and 1. The posterior distribution of $\theta_j$ is $$\theta_j | \bm{X} \sim Beta(\bm{b_i}(\bm{\alpha+X}),(\bm{1}-\bm{b_i})(\bm{\alpha+X})).$$ 

\subsection{Dual-criterion decision rule \label{decisionrule}}

The dual-criterion decision-making framework is based on the comparison of $\theta$ with multilevel target product profiles. Let $\theta_{LRV}$ be the lower reference value and $\theta_{CMV}$ be the clinically meaningful value. In single-arm trials, $\theta_{LRV}$ could denote the treatment effect of a historical control, while $\theta_{CMV}$ could denote the minimal clinically meaningful treatment effect. The value of $\theta_{LRV}$ and $\theta_{CMV}$ are prespecified and determined by trialists based on clinical considerations and market expectations. The new treatment is considered statistically effective if it shows high posterior probability that $\theta > \theta_{LRV}$, and it is considered clinically superior when it shows moderate posterior probability that $\theta > \theta_{CMV}$. The interim and final decisions are made based on these two criteria. In what follows, we first describe the decision rule at the end of the trial, followed by the decision rule at the interims. 

Let $\lambda_{_{LRV}}$ and $\lambda_{_{CMV}}$ denote probability cutoffs. At the end of the trial, we make one of three mutually exclusive decisions based on the following Bayesian posterior probability rule.  
\begin{itemize}
\item Go decision is made if  $P(\theta>\theta_{LRV}|data)>\lambda_{_{LRV}}$ and $P(\theta>\theta_{CMV}|data)>\lambda_{_{CMV}}$.
\item No-go decision is made if $P(\theta>\theta_{LRV}|data)<\lambda_{_{LRV}}$ and $P(\theta>\theta_{CMV}|data)<\lambda_{_{CMV}}$.
\item Consider decision is made otherwise.
\end{itemize}
Where go decision means that the drug is effective both statistically and clinically, and it warrants moving to the next stage of development; no-go decision means that the drug fails to show efficacy either statistically or clinically, and it should be terminated; and consider decision means that the drug shows efficacy either statistically or clinically, but not both, and the final decision whether the drug warrants further development should be made based on other considerations (e.g., other endpoints, and totality of the data). The procedure to determine the probabilities cutoffs $\lambda_{_{LRV}}$ and $\lambda_{_{CMV}}$ will be discussed in the next section.

Unlike at the end of the trial, the decision at interims is often binary: continue the study or stop for futility. Let $N$ denote the maximum sample size of the trial. We propose the following decision rule to make interim decisions.
\begin{itemize}
\item No-go decision is made if $P(\theta>\theta_{LRV}|data)<\lambda_{_{LRV}}(\frac{n}{N})^{\gamma_{_{LRV}}}$ and $P(\theta>\theta_{CMV}|data)<\lambda_{_{CMV}}(\frac{n}{N})^{\gamma_{_{CMV}}}$.
\item Continue decision is made otherwise.
\end{itemize}
Where ${\gamma_{_{LRV}}}$ and ${\gamma_{_{CMV}}}$ are tuning parameters that allow the probability cutoffs to adaptively vary with the fraction of information at the interim (i.e., $n/N$). Zhou et al. (2017) showed that such information-dependent cutoffs improve the operating characteristics of the design. 
The above decision rule says that we should stop the trial for futility if the drug fails to show efficacy either statistically or clinically; otherwise continue the trial to the next interim or the completion of the trial. 

Some remarks are warranted. First, the interim no-go decision rule is consistent and seamlessly merges with the no-go decision rule used at the end of the trial, noting that $n/N=1$ at the end of the trial. Second, we consider the binary decisions (continue/no-go) for interim, but it is not required by our design. If needed, the three decisions (go/consider/no-go) can also be used for interims. Lastly, we do not consider superiority stopping at the interim. The reason is that in phase II proof-of-concept trials, even if positive signals are observed, we often prefer continuing the study to collect more data, rather than stopping it. However, when appropriate, we can incorporate the superiority stopping by employing the three decisions rule described previously, but changing go decision to graduate decision (i.e., superiority stopping) and consider decision to continue decision.

The decision-making strategy for multiple endpoints is built upon that for each of the endpoints involved. In the following, we use ORR and EFS6 as an example of multiple endpoints to illustrate the strategy (see Figure 1).
\begin{enumerate}
\item At each of the interims, apply the interim decision rule described previously to ORR and EFS6 independently, and make the continue/no-go decision as follows.
\begin{itemize}
\item No-go decision is made if the decision for ORR and EFS6 are both no-go.
\item Continue decision is made otherwise.
\end{itemize}
\item At the end of the trial, apply the end-of-trial decision rule described previously to ORR and EFS6 independently, and make the go/consider/no-go decision as follows.
\begin{itemize}
\item Go decision is made if the decision for ORR or EFS6 is go.
\item No-go decision is made if the decisions for ORR and EFS6 are both no-go.
\item Consider decision is made otherwise.
\end{itemize}
\end{enumerate}

Similarly, the decision-making strategy for co-primary endpoints is also built upon that for each of the endpoints involved. We use toxicity rate (TR) and ORR as an example of co-primary endpoints to illustrate the strategy (see Figure 2), described as follows.
\begin{enumerate}
\item At each of interims, apply the interim decision rule to TR and ORR independently, and make trial continue/no-go decision as follows.
\begin{itemize}
\item No-go decision is made if the decisions for TR or ORR is no-go.
\item Continue decision is made otherwise.
\end{itemize}
\item At the end of the trial, apply the end-of-trial decision rule to TR and ORR independently, and make trial go/consider/no-go decision as follows.
\begin{itemize}
\item Go decision is made if the decisions for TR and ORR are both go.
\item No-go decision is made if the decision for TR or ORR is no-go.
\item Consider decision is made otherwise.
\end{itemize}
\end{enumerate}
The strategy of combining the decisions from individual endpoints is not unique, and it should be tailored to the application. For example, when the decision of ORR is consider and the decision of EFS6 is no-go, in some trials, we may prefer the decision of no-go, rather than consider.

\subsection{Optimization of design parameters \label{optimization}}
This section discusses how to determine the design parameters of BOP2-DC, i.e., $\lambda_{_{LRV}}$, $\lambda_{_{CMV}}$, $\gamma_{_{LRV}}$, and $\gamma_{_{CMV}}$. Let $\theta_{futile}$ denote the treatment effect deemed futile, and  $\theta_{eff}$ denote the target treatment effect. We define the following four performance metrics 
\begin{itemize}
\item False go rate (FGR) = $P(go|\theta=\theta_{futile})$ is defined as the probability of making a go decision when the treatment is futile.
\item False no-go rate (FNGR) = $P(no \,\, go|\theta=\theta_{eff})$ is defined as the probability of making a no-go decision when the treatment is effective.
\item Correct go rate (CGR) = $P(go|\theta=\theta_{eff})$ is defined as the probability of making a go decision when the treatment is effective.
\item False consider rate (FCR) = max\{$P(consider | \theta=\theta_{futile})$, $P(consider | \theta=\theta_{eff})$\} is defined as the probability of making a consider decision when the treatment is either futile or effective.

\end{itemize}
FGR, FNGR, and CGR are analogous to type I error, type II error,  and power in the hypothesis testing framework with binary decisions. As we here allow three decisions, unlike power = 1-type II error,  FNGR is not equal to 1-CGR in most cases. FCR is unique for the dual-criterion decision-making framework. In practice, it is desirable to set an upper limit for FCR because a consider decision after trials is not preferable. $\theta_{futile}$ and $\theta_{eff}$ should be prespecified based on clinical utility. In practice, $\theta_{futile}$ could be taken as $\theta_{_{LRV}}$, but not necessary, and $\theta_{eff}$ should be larger than or equal to $\theta_{_{CMV}}$.

We propose two algorithms to choose design parameters $\lambda_{_{LRV}}$, $\lambda_{_{CMV}}$, $\gamma_{_{LRV}}$, and $\gamma_{_{CMV}}$: (i) given the maximum sample size $N$, we maximize CGR while controlling FGR and FNGR at a prespecified levels, referred to as optimal design; and (ii) we minimize the expected sample size when experimental treatment is futile while controlling FGR and FNGR at a prespecified levels, referred to as minN design.  The optimization can be done using a grid search. In general, we may consider $\lambda_{_{LRV}} \in [0.5,0.99]$,  $\lambda_{_{CMV}} \in [0.01,0.5]$.  $\gamma_{_{LRV}}$ and $\gamma_{_{CMV}}$ are chosen within $[0,1]$ in order to avoid early decision boundaries that are too restrictive. Further clinical information may be helpful to refine the range of the search grid to speed up the search.

\subsection{Randomized controlled trials}
The proposed methodology is readily extended to randomized controlled trials (RCT). Let $\theta_E$ and $\theta_C$ denote the treatment effect (e.g., response rate, mean biomarker level, or median survival) for the experimental arm and the control arm, respectively. As subjects randomized to the treatment arm and the control arm are independent, the Bayesian models described previously can be applied to each of the arms independently to obtain the posterior distributions of $\theta_E$ and $\theta_C$. Define $\theta = \theta_E - \theta_C$ to represent the treatment effect improvement of the experimental arm on the control arm. Let $\theta_{LRV}$ be the lower reference value of $\theta$, representing the tolerance of treatment effect difference, and $\theta_{CMV}$ represent the minimal clinically meaningful treatment effect improvement. The Bayesian dual-criterion decision rule described in Section 2.2 can be directly used to make the interim ``continue/no-go" decisions and final ``go/consider/no-go" decisions based on the posterior probability of $\theta$. 

In some RCT, it may be desirable to consider both futility and superiority stopping at interims. In this case, we only need to modify the interim decision rule as follows.
\begin{itemize}
\item Graduate decision is made if  $P(\theta > \theta_{LRV}|data)>2\Phi(Z_{\frac{1+\lambda_{_{LRV}}}{2}}/\sqrt{\frac{n}{N}})-1$ and $P(\theta > \theta_{CMV}|data)>2\Phi(Z_{\frac{1+\lambda_{_{CMV}}}{2}}/\sqrt{\frac{n}{N}})-1$.
\item No-go decision is made if $P(\theta>\theta_{LRV}|data)<\lambda_{_{LRV}}(\frac{n}{N})^{\gamma_{_{LRV}}}$ and $P(\theta>\theta_{CMV}|data)<\lambda_{_{CMV}}(\frac{n}{N})^{\gamma_{_{CMV}}}$.
\item Continue decision is made otherwise.
\end{itemize}
Where $\Phi(\cdot)$ is the cumulative density function of the standard normal distribution, and $Z_q$ denotes the $q$ percentile of the standard normal distribution. The probability cutoffs in the graduate decision rule are chosen to be consistent with the widely used O’Brien-Fleming boundary. The same optimization procedure described in Section \ref{optimization} can be used to optimize the design parameters that appeared in the decision rule (e.g., $\lambda_{_{LRV}}$, $\lambda_{_{CMV}}$, $\gamma_{_{LRV}}$, and $\gamma_{_{CMV}}$). 

\section{Simulation}
\subsection{Single-arm trials}
\subsubsection{Setup}
We evaluated the operating characteristics of the BOP2-DC design using simulations. We performed 10,000 simulated trials for a binary endpoint,  continuous endpoint,  time-to-event endpoint, multiple endpoints,  and co-primary endpoints (i.e., toxicity and efficacy) with different scenarios. We assumed the maximum sample size in each simulated trial is $N=40$, and performed three interim analyses when the sample size reached 10, 20 and 30.  
We compared the operating characteristics of the BOP2-DC design with those of the original BOP2 design (Zhou et al., 2017), which was based on the binary decision framework. Of note, the primary objective of this comparison is to delineate the difference between the two frameworks, rather than showing that one framework is superior to the other. This is because it is difficult, if not impossible, to make fair comparison between the two designs, as the performance metrics for the binary-decision-based BOP2, including type I error and power, are not well defined in the dual-criteria-based (or trinary-decision-based) BOP2-DC. To make the comparison more interpretable, we calibrated the BOP2 design by controlling FGR (type I error) at 10\% and the BOP2-DC designs by controlling FGR at 5\%, FNGR at 10\% and setting the upper limit of FCR as 20\%. The reason we used a lower FGR for BOP2-DC is that in the BOP2 design, a false go decision is the only incorrect decision that could be made under the null hypothesis, while under the BOP2-DC designs, incorrect decisions under the null hypothesis can be caused by either false go decision or false consider decision.  

\subsubsection{Results}
Table 1 shows the operating characteristics of BOP2-DC (including optimal and minN designs) and BOP2 in a single-arm trial for a binary/continuous/time-to-event endpoint. In scenarios 1-3, we assume $\theta_{futile}=0.2$, $\theta_{eff}=0.4$, $\theta_{_{LRV}}=0.2$, and $\theta_{_{CMV}}=0.3$. Scenario 1 represents the case in which the treatment is not effective. The BOP2 design yields 7.4\% FGR and the BOP2-DC designs yield around 4\% FGR. The FGR of all three designs is lower than the nominal value due to the discrete nature of the binary endpoint. The minN design has the smallest average sample size when treatment is not effective compared to the other two designs.  Scenario 2 represents the case that the drug does not achieve the minimal clinically meaningful effect. Both optimal and minN designs yield around 30\% FGR compared to 39.7\% in the BOP2 design, and the probability to make a ``consider" decision in the optimal and minN designs is 34.0\% and 22.9\%, respectively. This gives researchers the flexibility to incorporate other clinical considerations when the treatment effect is close to the minimal clinically meaningful value. Scenario 3 represents the case that the drug has an effective treatment effect. The optimal and minN design respectively have CGR at 85.9\% and 82.7\%, and FNGR at 3.6\% and 9.7\%, while the BOP2 design has CGR at 88.6\% and FNGR at 11.4\%. Although CGR in the BOP2-DC designs are slightly lower than that in BOP2, FNGR is much lower in the BOP2-DC designs, which means that if a treatment is effective, the BOP2-DC designs would have lower probability to (incorrectly) claim it futile.

In scenarios 4-6, we assume $\theta_{futile}=0$, $\theta_{eff}=0.2$ for a single continuous endpoint, with $\theta_{_{LRV}}=0$ and $\theta_{_{CMV}}=0.1$. Scenario 4 represents the case that treatment is ineffective. The BOP2 design yields 9.5\% FGR,  while both the optimal and minN designs control FGR under 5\%. Scenario 5 represents the case that treatment does not achieve the minimal clinically meaningful effect. Both the optimal and minN designs yield  lower FGR (25.1\% and 23.6\%, respectively) compared to 37.6\% in BOP2, and a high probability (35.1\% and 31.7\%) to make a consider decision. Scenario 6 represents the case that treatment is effective. Similar to the results in Scenario 3, CGR of the optimal and minN designs are lower than that of BOP2, but FNGR is also much lower in the BOP2-DC designs. 

In scenarios 7-9, the endpoint is the time to event (e.g., PFS). We assume $\theta_{futile}=6$ months, $\theta_{eff}=10$ months for the median PFS, with $\theta_{LRV}=6$ months and $\theta_{CMV}=8$ months. We assume that the accrual rate is 1 month and the follow-up time after the last patient is 12 months.  Scenarios 7-9 represent the cases that patients after having received treatment, have different levels of median survival time from the lowest to the highest. Simulation results are similar to these in scenarios 1-6. BOP2 controls FGR at 10\%, and the optimal and minN designs control FGR at 5\% in Scenario 7. In Scenario 8, the optimal and minN designs yield high FCR at 27.7\% and 17.5\%. In Scenario 9, all three designs yield comparable CGR, while the optimal and minN designs each have a much lower FNGR than BOP2.

Table 2 shows the operating characteristics for multiple endpoints in single-arm trials. We use $\theta^{(i)}_{futile}$ and $\theta^{(i)}_{eff}, i=1,2$ to denote the futile treatment effect and effective treatment effect for two endpoints, and use $\theta^{(i)}_{_{LRV}}$ and $\theta^{(i)}_{_{CMV}}, i=1,2$ to represent the lower reference value and clinically meaningful value for two endpoints. We assume $\theta^{(1)}_{futile}=0.15$, $\theta^{(1)}_{eff}=0.25$, $\theta^{(1)}_{LRV}=0.15$ and $\theta^{(1)}_{CMV}=0.2$ for the first endpoint, and  $\theta^{(2)}_{futile}=0.2$, $\theta^{(2)}_{eff}=0.4$, $\theta^{(2)}_{LRV}=0.2$, and $\theta^{(2)}_{CMV}=0.3$ for the second endpoint. In Scenario 1, BOP2 controls FGR at 10\% and the optimal and minN designs control FGR at 5\%. All three designs yield correct no-go rates around 90\% when treatment is ineffective. In Scenario 2, the drug is not sufficiently effective, and BOP2-DC have a smaller FGR and comparable correct no-go rate, compared to BOP2. In Scenario 3 and 4, the treatment is effective. The proposed optimal design has a slightly lower CGR compared to BOP2, but the overall operating characteristics are similar to BOP2 between the two designs. The minN design is the most conservative among the three designs, as it has a much lower go decision rate and average sample size in all scenarios. 

Table 3 shows the operating characteristics for efficacy and toxicity endpoints in a single-arm trial. Both efficacy and toxicity endpoints are binary. For the efficacy endpoint, we assume $\theta_{futile}=0.3$, $\theta_{eff}=0.5$. For the toxicity endpoint, we assume $\theta_{toxic}=0.2$, $\theta_{safe}=0.1$, where $\theta_{toxic}$ is the toxicity rate regarded as unacceptable, and $\theta_{safe}$ is the toxicity rate regarded as safe.  We use $(\theta_{_{LRV, EFF}}, \theta_{_{LRV, TOX}}) = (0.3,0.2)$ and $(\theta_{_{CMV, EFF}}, \theta_{_{CMV, TOX}}) = (0.45,0.15)$ to represent the lower reference value and clinically meaningful value for efficacy and toxicity endpoints. In Scenario 1, when treatment is futile and toxic, BOP2 controls FGR at 10\% and the optimal and minN designs control FGR at 5\%. In Scenario 2, when the treatment is ineffective and safe, the optimal and minN designs have smaller FGR (25.3\% and 22.7\%) and comparable correct no-go rate (40.1\% and 52.6\%) compared to 43.6\% FGR and 56.4\% correct no-go rate in BOP2. In Scenario 3, when treatment is effective and safe, the proposed optimal and minN designs have comparable CGR (90.8\% and 86.4\%) and lower FNGR (4.2\% and 10.0\%), compared to 88.7\% CGR and 11.3\% FNGR in the BOP2 design.\\

\subsection{Randomized trials}
We also evaluated the operating characteristics of the BOP2-DC designs with randomized controlled trials using simulations. We performed 2000 simulated trials with different scenarios. We assumed the maximum sample size in simulated trials was equal to $N=75$, and performed three interim analyses when sample size reaches 30, 45 and 60 for a single binary endpoint. Within each interim analysis and final analysis, we assume that patients are assigned into the experimental arm and the control arm with a ratio of 2:1. While the assignment ratio would not necessarily be 2:1 in the proposed design, it could be chosen by the need of clinicians. Similar to single arm trial designs, the optimization algorithms calibrate the threshold parameters $\lambda_{_{LRV}}$, $\lambda_{_{CMV}}$, $\gamma_{_{LRV}}$, and $\gamma_{_{CMV}}$ while controlling FGR at 5\%, FNGR at 30\%, and FCR smaller than 20\%.

Table 4 shows the operating characteristics for a binary endpoint (e.g., response rate) in a randomized trial. We assume the response rate of the control arm is 0.2, i.e., $\theta_{C}=0.2$,  and the experimental arm is considered as futile when $\theta_{E}=0.2$ and effective when $\theta_{E}=0.4$. We set $\theta_{LRV}=0$ and $\theta_{CMV}=0.2$. In Scenario 1, when treatment is futile, BOP2-DC designs control FGR at 5\% and BOP2 controls FGR at 10\%. In Scenario 2, the treatment in the experimental arm is not sufficiently effective compared to the control arm, the optimal and minN designs each have a smaller FGR and smaller correct no-go rate compared to BOP2. In Scenario 3, when the experimental arm is effective. The optimal design has a slightly smaller CGR and FNGR compared to BOP2. Table S1-S4 in Supplementary Materials show the operating characteristics for a single continuous endpoint, single time-to-event endpoint, multiple endpoints and efficacy and toxicity endpoints in a randomized trial, respectively. The results are generally similar to those above. That is, the optimal and minN designs control FGR and FNGR at desirable levels when treatment is either effective or ineffective, and they yield a high FCR when treatment is not sufficiently effective. That provides researchers the flexibility to make further decision by accounting for other data information (e.g., secondary endpoints). \\

\section{Discussion}
We have proposed the BOP2-DC designs to incorporate statistical significance and clinical relevance into decision makings for phase II trials. By employing the Bayesian framework, BOP2-DC is highly flexible and accommodates various types of endpoints, including binary, continuous, time-to-event, multiple, and co-primary endpoints, for both single and two-arm trials. BOP2-DC is efficient as it maximizes the probability of a go decision when the treatment is effective, or it minimizes the average sample size when the treatment is ineffective. Simulation shows that BOP2-DC has desirable operating characteristics.

Depending on the application, BOP2-DC can be modified to better fit trial objectives. For example, BOP2-DC controls FGR, FNGR and FCR, and it optimizes the probability of a go decision when the treatment is effective. In some cases, clinicians may be interested in controlling FGR+FCR (the probability of not making a go decision when treatment is futile) or FNGR+FCR (the probability of not making a no-go decision when treatment is effective). This can be done by modifying the optimization algorithm of BOP2-DC. Another extension of BOP2-DC is to extend it to basket trials. The basic idea of BOP2-DC is readily applicable, but a new modeling strategy should be employed to estimate the treatment effect of each arm, with the potential to borrow information across the arms, e.g., using Bayesian hierarchical model (Thall et al., 2003; Berry et al., 2013, Chu and Yuan, 2018) or more efficient clustered Bayesian hierarchical model (Jiang et al., 2021).
\\

\pagebreak

\newpage

\begin{figure}
\includegraphics[width=0.8\textwidth]{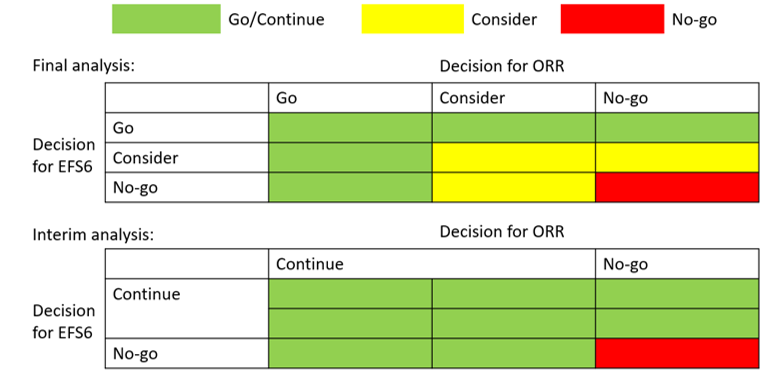}
\caption{Decision rule for multiple endpoints}
\end{figure}

\begin{figure}
\includegraphics[width=0.8\textwidth]{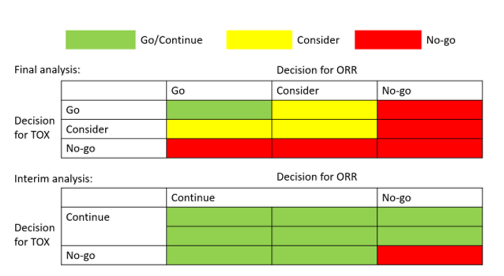}
\caption{Decision rule for co-primary endpoints (e.g. toxicity and efficacy)}
\end{figure}
 
\begin{table}
\caption{Go rate, no-go rate, consider rate and average sample size for the proposed single arm BOP2-DC designs (optimal and minN) and the BOP2 design for a single binary endpoint, continuous endpoint and time-to-event endpoint.}
\begin{minipage}[b]{1.0\linewidth}
\begin{tabular}{p{0.4cm}p{2cm}p{1.4cm}p{0.9cm}p{1.3cm}p{1.7cm}p{1.5cm}p{1.5cm}p{2cm}} 
 \hline\hline
Sc. & Design & $\theta_{_{LRV}}$ & $\theta_{_{CMV}}$&  $\theta_{true}$ & Go rate (\%) & No-Go rate (\%) & Consider rate (\%) & Average sample size \\
\hline\hline
\end{tabular}
\end{minipage}

\begin{minipage}[b]{1.0\linewidth}
\begin{threeparttable}
\subcaption*{Binary endpoint: $\theta_{futile}=0.2$, $\theta_{eff}=0.4$}
 \begin{tabular}{p{0.4cm}p{2cm}p{1.4cm}p{0.9cm}p{1.3cm}p{1.7cm}p{1.5cm}p{1.5cm}p{2cm}}

\hline
$1^*$ & BOP2 & 0.2 & 0.3 & 0.2 & 7.4 & 92.6 &  0 & 23.2\\
& optimal& 0.2 & 0.3 & 0.2 & 4.3 & 77.6 & 18.1 & 28.3\\ 
  & minN& 0.2 & 0.3 & 0.2 & 3.7 & 84.4 & 11.9 & 21.5\\ 
\cline{1-9}
2 & BOP2 & 0.2 & 0.3 & 0.28 & 39.7 & 60.3 & 0  & 31.5\\
& optimal& 0.2 & 0.3 & 0.28 & 30.5 & 35.5 & 34.0 & 35.3\\
  & minN& 0.2 & 0.3 & 0.28 & 28.0 & 49.1 & 22.9 & 29.9\\
\cline{1-9}
$3^\dagger$ & BOP2 & 0.2 & 0.3 & 0.4 & 88.6 & 11.4 & 0  & 38.2\\
& optimal& 0.2 & 0.3 & 0.4 & 85.9 & 3.6 & 10.5 & 39.4\\
  & minN& 0.2 & 0.3 & 0.4 &82.7 & 9.7 & 7.6 & 37.8\\
\hline
\end{tabular}
\end{threeparttable}
\end{minipage}
\par
\begin{minipage}[b]{1.0\linewidth}
\begin{threeparttable}
\subcaption*{Continuous endpoint: $\theta_{futile}=0$, $\theta_{eff}=0.2$}
 \begin{tabular}{p{0.4cm}p{2cm}p{1.4cm}p{0.9cm}p{1.3cm}p{1.7cm}p{1.5cm}p{1.5cm}p{2cm}} 
 \hline
$ 4^*$ & BOP2 & 0 & 0.1 & 0 & 9.5 & 90.5 &0 & 26.0\\ 
  & optimal& 0 & 0.1 & 0  & 4.9 & 76.5 & 18.6 & 27.7\\ 
  & minN&0 & 0.1 & 0 & 4.5 & 77.7 & 17.9 & 22.7\\ 
\cline{1-9}
5 & BOP2 & 0 & 0.1 & 0.08 & 37.6 & 62.4 & 0 & 33.1\\
  & optimal& 0 & 0.1 & 0.08 & 25.1 & 39.8 & 35.1 & 34.4\\
  & minN& 0 & 0.1 & 0.08 & 23.6 & 44.7 & 31.7 & 30.2\\
\cline{1-9}
$ 6^\dagger$ & BOP2 & 0 & 0.1 & 0.2 & 87.9 & 12.1 & 0.0 & 38.9\\
  & optimal& 0 & 0.1 & 0.2 & 80.0 & 4.9 & 15.1 & 39.2\\
  & minN& 0 & 0.1 & 0.2 & 77.0 & 10.0 & 13.0 & 37.4\\
\hline
\end{tabular}
\end{threeparttable}
\end{minipage}
\par
\begin{minipage}[b]{1.0\linewidth}
\begin{threeparttable}
\subcaption*{Time-to-event endpoint: $\theta_{futile}=6$, $\theta_{eff}=10$}
 \begin{tabular}{p{0.4cm}p{2cm}p{1.4cm}p{0.9cm}p{1.3cm}p{1.7cm}p{1.5cm}p{1.5cm}p{2cm}} 
\hline
 $ 7^*$ & BOP2 & 6 & 8 & 6 & 9.9 & 90.1 & 0 & 22.5\\
 & optimal& 6 & 8 & 6  & 5.0 & 80.5 & 14.5 & 28.6\\
  & minN&6 & 8 & 6 & 4.6 & 87.8 & 7.6 & 24.8\\
\cline{1-9}
8& BOP2 & 6 & 8 & 7.5 & 45.5 & 54.5 & 0 & 29.9\\
& optimal& 6 & 8 & 7.5 & 37.8 & 34.5 & 27.7 & 35.2\\
  & minN& 6 & 8 & 7.5 & 35.0 & 47.5 & 17.5 & 32.2\\
\cline{1-9}
$ 9^\dagger$& BOP2 & 6 & 8 & 10 & 84.7 & 15.3 & 0 & 36.2\\
& optimal& 6 & 8 & 10 & 90.0 & 5.3 & 4.7 & 38.9\\
  & minN& 6 & 8 & 10 & 86.7 & 10.0& 3.3 & 37.8\\

  \hline
\end{tabular}
 \begin{tablenotes}
	 \item[*] Treatment is futile (null hypothesis)
     \item[$\dagger$] Treatment is effective (alternative hypothesis)
   \end{tablenotes}
\end{threeparttable}
\end{minipage}
\end{table}

\newpage

 \setcounter{table}{1}
\begin{table}
\begin{threeparttable}
\caption{Go rate, no-go rate, consider rate and average sample size for the proposed single arm BOP2-DC designs (optimal and minN) and the BOP2 design for multiple endpoints. $\theta^{(1)}_{futile}=0.15, \theta^{(2)}_{futile}=0.2, \theta^{(1)}_{eff}=0.25, \theta^{(2)}_{eff}=0.4$.}
\begin{center}
 \begin{tabular}{p{0.4cm}p{1.2cm}p{1.5cm}p{1.5cm}p{1.7cm}p{1.5cm}p{1.5cm}p{1.5cm}p{2cm}} 
 \hline
Sc. & Design & $\theta_{_{LRV}}$ & $\theta_{_{CMV}}$&  $\theta_{true}$ & Go rate (\%) & No-Go rate (\%) & Consider rate (\%) & Average sample size \\
\hline
$ 1^*$ & BOP2 & 0.15, 0.2 & 0.2, 0.3 & 0.15, 0.2 & 8.9 & 91.1 & 0.0 & 25.8\\ 
 & optimal&  0.15, 0.2 & 0.2, 0.3 & 0.15, 0.2 & 4.8 & 88.4 & 6.8 & 22.0\\ 
  & minN&  0.15, 0.2 & 0.2, 0.3 & 0.15, 0.2 & 3.3 & 91.9 & 4.8 & 17.3\\ 
\cline{1-9}
 2 & BOP2 & 0.15, 0.2 & 0.2, 0.3 & 0.18, 0.25 & 25.6 & 74.4 & 0.0 & 30.5\\ 
 & optimal&  0.15, 0.2 & 0.2, 0.3 & 0.18, 0.25  & 18.7& 72.6& 8.7 & 26.4\\
  & minN&  0.15, 0.2 & 0.2, 0.3 & 0.18, 0.25  &  14.4 & 80.0& 5.5 & 21.2\\
\cline{1-9}
 3 & BOP2 & 0.15, 0.2 & 0.2, 0.3 & 0.15, 0.4 & 83.2 & 16.8 & 0.0 & 38.0\\ 
 & optimal&  0.15, 0.2 & 0.2, 0.3 & 0.15, 0.4  & 75.1 & 24.2& 0.7 & 35.2\\
  & minN&  0.15, 0.2 & 0.2, 0.3 & 0.15, 0.4  & 66.7 & 33.0 & 0.3 & 32.6\\
\cline{1-9}
4$^\dagger$ & BOP2 & 0.15, 0.2 & 0.2, 0.3 & 0.25, 0.4& 92.3 & 7.7 & 0.0 & 39.3\\ 
& optimal&  0.15, 0.2 & 0.2, 0.3 & 0.25, 0.4 & 89.0 & 9.8 & 1.2 & 38.2\\ 
  & minN&  0.15, 0.2 & 0.2, 0.3 & 0.25, 0.4 & 81.2 & 18.1 & 0.7 & 35.9\\ 

\hline
\end{tabular}
 \begin{tablenotes}
	 \item[*] Treatment is futile (null hypothesis)
     \item[$\dagger$] Treatment is effective (alternative hypothesis)
   \end{tablenotes}
\end{center}
\end{threeparttable}
\end{table}

\begin{table}
\begin{threeparttable}
\caption{Go rate, no-go rate, consider rate and average sample size for the proposed single arm BOP2-DC designs (optimal and minN) and the BOP2 design for efficacy and toxicity endpoints. For efficacy endpoint, we assume $\theta_{futile}=0.3$, $\theta_{eff}=0.5$. For toxicity endpoint, we assume $\theta_{toxic}=0.2$, $\theta_{safe}=0.1$.}
\begin{center}
 \begin{tabular}{p{0.4cm}p{1.2cm}p{1.5cm}p{1.7cm}p{1.7cm}p{1.5cm}p{1.5cm}p{1.5cm}p{2.2cm}} 
 \hline
Sc. & Design & $\theta_{_{LRV}}$ & $\theta_{_{CMV}}$&  $\theta_{true}$ & Go rate (\%) & No-Go rate (\%) & Consider rate (\%) & Average sample size \\
\hline
$ 1^*$  & BOP2 & 0.3, 0.2 & 0.4, 0.15 & 0.3, 0.2 & 9.6& 90.4 & 0 & 22.9 \\ 
& optimal&  0.3, 0.2 & 0.4, 0.15 & 0.3, 0.2 & 4.7 & 78.5 & 16.8 & 27.7\\ 
  & minN&  0.3, 0.2 & 0.4, 0.15 & 0.3, 0.2  & 4.3 & 84.2 & 11.5 & 21.7 \\ 
 \cline{1-9}
$ 2$  & BOP2 & 0.3, 0.2 & 0.4, 0.15 & 0.35, 0.1 & 43.6& 56.4 & 0 & 31.3 \\ 
& optimal&  0.3, 0.2 & 0.4, 0.15 & 0.35, 0.1 & 25.3 & 40.1 & 34.6 & 34.5 \\ 
  & minN&  0.3, 0.2 & 0.4, 0.15 & 0.35, 0.1  & 22.7 & 52.6 & 24.7 & 28.8\\ 
\cline{1-9}
$ 3^\dagger$  & BOP2 & 0.3, 0.2 & 0.4, 0.15 & 0.5, 0.1 & 88.7& 11.3 & 0 & 38.0 \\ 
& optimal&  0.3, 0.2 & 0.4, 0.15 & 0.5, 0.1 & 90.8 & 4.2 & 5.0 & 39.3\\ 
  & minN&  0.3, 0.2 & 0.4, 0.15 & 0.5, 0.1  & 86.4 & 10.0 & 3.6 & 37.6 \\ 

\hline
\end{tabular}
 \begin{tablenotes}
	 \item[*] Treatment is futile and toxic (null hypothesis)
     \item[$\dagger$] Treatment is effective and safe (alternative hypothesis)
   \end{tablenotes}
\end{center}
\end{threeparttable}
\end{table}

\begin{table}
\begin{threeparttable}
\caption{Go rate, no-go rate, consider rate and average sample size for the proposed two-arm BOP2-DC designs (optimal and minN) and the BOP2 design for a single binary endpoint. Experimental treatment is considered as futile when $\theta_{E}=\theta_{C}=0.2$ and experimental treatment is considered as effective when $\theta_{E}=0.4$.}
\begin{center}
 \begin{tabular}{p{0.4cm}p{2cm}p{1cm}p{1cm}p{1cm}p{1cm}p{1.5cm}p{1.5cm}p{1.5cm}p{2.2cm}} 
 \hline
Sc. & Design & $\theta_{LRV}$ & $\theta_{CMV}$&  $\theta_{true,C}$&  $\theta_{true,E}$ & Go rate (\%) & No-Go rate (\%) & Consider rate (\%) & Average sample size \\
\hline
$1^*$ & BOP2 & 0 & 0.2 &0.2 & 0.2 & 9.4 & 90.6 & 0 & 49.1\\ 
& optimal & 0 & 0.2 &0.2 & 0.2  & 4.7& 81.4 & 13.9 & 51.3 \\ 
  & minN &0 & 0.2 &0.2 & 0.2 & 4.9 & 83.4 & 11.7 & 40.0\\ 
\cline{1-10}
2 & BOP2 & 0 & 0.2 &0.2 & 0.28 & 27.3 & 72.7 & 0 & 55.4\\ 
& optimal& 0 &0.2 &0.2& 0.28  & 19.2& 58.7  & 22.2 & 56.3 \\
  & minN & 0 &0.2 &0.2& 0.28  & 18.7 & 62.7 & 18.6 & 43.7\\
\cline{1-10}
$3^\dagger$& BOP2 & 0 & 0.2 &0.2 & 0.4 & 65.2 & 34.8 & 0 & 56.6\\ 
& optimal& 0 &0.2 &0.2& 0.4  &  61.4 & 21.7 & 16.9 & 53.1\\
  & minN & 0&0.2 &0.2& 0.4 & 56.5 & 29.8 & 13.7 & 44.0\\
 
\hline
\end{tabular}
 \begin{tablenotes}
	 \item[*] Experimental arm is futile (null hypothesis)
     \item[$\dagger$] Experimental arm is effective (alternative hypothesis)
   \end{tablenotes}
\end{center}
\end{threeparttable}
\end{table}


\newpage
\begin{thebibliography}{9}



\bibitem[Berry et al., 2013]{Berry}\label{Berry}
Berry SM, Broglio KR, Groshen S, et al. Bayesian hierarchical modeling of patient subpopulations: efficient designs of phase II oncology clinical trials. \textit{Clinical Trials} 2013; 10(5), 720-734. doi: 10.1177/1740774513497539.

\bibitem{Bertsche}  Bertsche, A., Fleischer, F., Beyersmann, J., \& Nehmiz, G. (2019). Bayesian Phase II optimization for time-to-event data based on historical information. \textit{Statistical Methods in Medical Research};28(4):1272-1289. doi: 10.1177/0962280217747310.

\bibitem[Chu and Yuan, 2018a]{Chu}\label{Chu}
Chu Y and Yuan Y. A Bayesian basket trial design using a calibrated Bayesian hierarchical model. \textit{Clinical Trials} 2018; 15(2), 149-158. doi: 10.1177/1740774518755122. 

\bibitem{Dunyak} Dunyak, J., Mitchell, P., Hamrén, B., Helmlinger, G., Matcham, J., Stanski, D., \& Al‐Huniti, N. (2018). Integrating dose estimation into a decision‐making framework for model‐based drug development. \textit{Pharmaceutical Statistics};17(2):155-168. doi: 10.1002/pst.1841.

\bibitem{Fisch} Fisch, R., Jones, I., Jones, J., Kerman, J., Rosenkranz, G. K., \& Schmidli, H. (2015). Bayesian design of proof-of-concept trials. \textit{Therapeutic Innovation \& Regulatory Science};49(1):155-162. doi: 10.1177/2168479014533970.

\bibitem{Frewer} Frewer, P., Mitchell, P., Watkins, C., \& Matcham, J. (2016). Decision‐making in early clinical drug development. \textit{Pharmaceutical Statistics};15(3):255-63. doi: 10.1002/pst.1746.

\bibitem{Jiang} Jiang L, Li R, Yan F, Yap TA, Yuan Y. Shotgun: A Bayesian seamless phase I-II design to accelerate the development of targeted therapies and immunotherapy. \textit{Contemporary Clinical Trials}. 2021 May;104:106338. doi: 10.1016/j.cct.2021.106338.

\bibitem{Kirby} Kirby, S., \& Chuang‐Stein, C. (2017). A comparison of five approaches to decision‐making for a first clinical trial of efficacy. \textit{Pharmaceutical Statistics};16(1):37-44. doi: 10.1002/pst.1775.

\bibitem{Lalonde} Lalonde, R. L., Kowalski, K. G., Hutmacher, M. M., Ewy, W., Nichols, D. J., Milligan, P. A., ... \& Miller, R. (2007). Model‐based drug development. \textit{Clinical Pharmacology \& Therapeutics};82(1):21-32. doi: 10.1038/sj.clpt.6100235.

\bibitem{Mander} Mander, A. P., Wason, J. M., Sweeting, M. J., \& Thompson, S. G. (2012). Admissible two‐stage designs for phase II cancer clinical trials that incorporate the expected sample size under the alternative hypothesis. \textit{Pharmaceutical Statistics};11(2):91-6. doi: 10.1002/pst.501.


\bibitem{Moore}  Moore KN, Sill MW, Tenney ME, Darus CJ, Griffin D, Werner TL, Rose PG, \&  (2015) Behrens R. A phase II trial of trebananib (AMG 386; IND\#111071), a selective angiopoietin 1/2 neutralizing peptibody, in patients with persistent/recurrent carcinoma of the endometrium: An NRG/Gynecologic Oncology Group trial. \textit{Gynecologic Oncology}; 138(3):513-8. doi: 10.1016/j.ygyno.2015.07.006.

\bibitem{} Pulkstenis, E., Patra, K., \& Zhang, J. (2017). A bayesian paradigm for decision-making in proof-of-concept trials. \textit{Journal of Biopharmaceutical Statistics}:442-456. doi: 10.1080/10543406.2017.1289947.

\bibitem{} Quan, H., Chen, X., Lan, Y., Luo, X., Kubiak, R., Bonnet, N., \& Paux, G. (2020). Applications of Bayesian analysis to proof‐of‐concept trial planning and decision making. \textit{Pharmaceutical Statistics};19(4):468-481. doi: 10.1002/pst.1985.

\bibitem{} Roychoudhury, S., Scheuer, N., \& Neuenschwander, B. (2018). Beyond p-values: a phase II dual-criterion design with statistical significance and clinical relevance. \textit{Clinical Trials};15(5):452-461. doi: 10.1177/1740774518770661.

\bibitem{} Sargent, D. J., Chan, V., \& Goldberg, R. M. (2001). A three-outcome design for phase II clinical trials. \textit{Controlled Clinical Trials};22(2), 117-125. doi: 10.1016/s0197-2456(00)00115-x.

\bibitem[Thall et al., 2003]{Thall}\label{Thall}
Thall PF, Wathen JK, Bekele BN, et al. Hierarchical Bayesian approaches to phase II trials in diseases with multiple subtypes. \textit{Statistics in Medicine} 2003; 22(5), 763-780. doi: 10.1002/sim.1399.

\bibitem{} Thall, P. F., Wooten, L. H., \& Tannir, N. M. (2005). Monitoring event times in early phase clinical trials: some practical issues. \textit{Clinical Trials}; 2(6):467-478.doi: 10.1191/1740774505cn121oa

\bibitem{Walley} Walley, R. J., Smith, C. L., Gale, J. D., \& Woodward, P. (2015). Advantages of a wholly Bayesian approach to assessing efficacy in early drug development: a case study. \textit{Pharmaceutical Statistics};14(3):205-15. doi: 10.1002/pst.1675.

\bibitem{} Zhou, H., Chen, C., Sun, L., \& Yuan, Y. (2020). Bayesian optimal phase II clinical trial design with time‐to‐event endpoint. \textit{Pharmaceutical Statistics};19(6):776-786. doi: 10.1002/pst.2030

\end{thebibliography}
\end{document}


\def\eqx"#1"{{\label{#1}}}
\def\eqn"#1"{{\ref{#1}}}

\makeatletter 
\@addtoreset{equation}{section}
\makeatother  

\def\yincomment#1{\vskip 2mm\boxit{\vskip 2mm{\color{red}\bf#1} {\color{blue}\bf --Yin\vskip 2mm}}\vskip 2mm}
\def\lincomment#1{\vskip 2mm\boxit{\vskip 2mm{\color{blue}\bf#1} {\color{black}\bf --Lin\vskip 2mm}}\vskip 2mm}
\def\squarebox#1{\hbox to #1{\hfill\vbox to #1{\vfill}}}
\def\boxit#1{\vbox{\hrule\hbox{\vrule\kern6pt
          \vbox{\kern6pt#1\kern6pt}\kern6pt\vrule}\hrule}}

\def\theequation{\thesection.\arabic{equation}}
\newcommand{\ds}{\displaystyle}

\newcommand{\bJ}{\mbox{\bf J}}
\newcommand{\bF}{\mbox{\bf F}}
\newcommand{\bM}{\mbox{\bf M}}
\newcommand{\bR}{\mbox{\bf R}}
\newcommand{\bZ}{\mbox{\bf Z}}
\newcommand{\bX}{\mbox{\bf X}}
\newcommand{\bx}{\mbox{\bf x}}
\newcommand{\bww}{\mbox{\bf w}}
\newcommand{\bQ}{\mbox{\bf Q}}
\newcommand{\bH}{\mbox{\bf H}}
\newcommand{\bh}{\mbox{\bf h}}
\newcommand{\bz}{\mbox{\bf z}}
\newcommand{\br}{\mbox{\bf r}}
\newcommand{\ba}{\mbox{\bf a}}
\newcommand{\be}{\mbox{\bf e}}
\newcommand{\bG}{\mbox{\bf G}}
\newcommand{\bB}{\mbox{\bf B}}
\newcommand{\bb}{\mbox{\bf b}}
\newcommand{\bA}{\mbox{\bf A}}
\newcommand{\bC}{\mbox{\bf C}}
\newcommand{\bI}{\mbox{\bf I}}
\newcommand{\bD}{\mbox{\bf D}}
\newcommand{\bU}{\mbox{\bf U}}
\newcommand{\bc}{\mbox{\bf c}}
\newcommand{\bd}{\mbox{\bf d}}
\newcommand{\bs}{\mbox{\bf s}}
\newcommand{\bS}{\mbox{\bf S}}
\newcommand{\bV}{\mbox{\bf V}}
\newcommand{\bv}{\mbox{\bf v}}
\newcommand{\bW}{\mbox{\bf W}}
\newcommand{\bY}{\mathbf{ Y}}
\newcommand{\bw}{\mbox{\bf w}}
\newcommand{\bg}{\mbox{\bf g}}
\newcommand{\bu}{\mbox{\bf u}}
\newcommand{\mI}{\mbox{I}}

\newcommand{\bch}{\color{blue}\it}
\newcommand{\ech}{\color{black}\rm}

\def\bb{{\bf b}}

\newcommand{\bcU}{\boldsymbol{\cal U}}
\newcommand{\bbeta}{\boldsymbol{\beta}}
\newcommand{\bdelta}{\boldsymbol{\delta}}
\newcommand{\bDelta}{\boldsymbol{\Delta}}
\newcommand{\boldeta}{\boldsymbol{\eta}}
\newcommand{\bxi}{\boldsymbol{\xi}}
\newcommand{\bGamma}{\boldsymbol{\Gamma}}
\newcommand{\bSigma}{\boldsymbol{\Sigma}}
\newcommand{\balpha}{\boldsymbol{\alpha}}
\newcommand{\bOmega}{\boldsymbol{\Omega}}
\newcommand{\btheta}{\boldsymbol{\theta}}
\newcommand{\bepsilon}{\boldsymbol{\epsilon}}
\newcommand{\bmu}{\boldsymbol{\mu}}
\newcommand{\bnu}{\boldsymbol{\nu}}
\newcommand{\bgamma}{\boldsymbol{\gamma}}
\newcommand{\btau}{\boldsymbol{\tau}}
\newcommand{\bTheta}{\boldsymbol{\Theta}}

\newtheorem{thm}{Theorem}
\newtheorem{lem}{Lemma}[section]
\newtheorem{rem}{Remark}[section]
\newtheorem{cor}{Corollary}[section]
\newcolumntype{L}[1]{>{\raggedright\let\newline\\\arraybackslash\hspace{0pt}}m{#1}}
\newcolumntype{C}[1]{>{\centering\let\newline\\\arraybackslash\hspace{0pt}}m{#1}}
\newcolumntype{R}[1]{>{\raggedleft\let\newline\\\arraybackslash\hspace{0pt}}m{#1}}

\newcommand{\beginsupplement}{%
        \setcounter{table}{0}
        \renewcommand{\thetable}{S\arabic{table}}%
        \setcounter{figure}{0}
        \renewcommand{\thefigure}{S\arabic{figure}}%
     }

\section*{Appendix}

\beginsupplement
\begin{table}[htp]
\begin{threeparttable}
\caption{Go rate, no-go rate, consider rate and average sample size for the proposed two-arm BOP2-DC designs (optimal and minN) and the BOP2 design for single continuous endpoint. Experimental treatment is considered as futile when $\theta_{E}=\theta_{C}=0$ and experimental treatment is considered as effective when $\theta_{E}=0.7$. Maximum sample size is 60, interim analysis is conducted when sample size reaches 30, 45.}
\begin{center}
 \begin{tabular}{p{0.4cm}p{1.7cm}p{0.7cm}p{0.7cm}p{0.7cm}p{0.7cm}p{1.5cm}p{1.5cm}p{1.5cm}p{2cm}} 
 \hline
Sc. & Design & $\theta_{_{LRV}}$ & $\theta_{_{CMV}}$&  $\theta_{C}$&  $\theta_{E}$ & Go rate (\%) & No-go rate (\%) & Consider rate (\%) & Average sample size \\
\hline
$1^*$ & BOP2 & 0 & 0.35 &0 & 0  & 9.8 & 90.2 & 0 & 43.0 \\ 
& optimal& 0 & 0.35 &0 & 0  & 5.0 & 88.3 & 6.4 & 38.1 \\ 
  & minN&0 & 0.35 &0 & 0 & 4.5 & 93.5 & 2.0 & 34.8 \\ 
\cline{1-10}
2  & BOP2 & 0 & 0.35 &0 & 0.3  & 42.7 & 57.3 & 0 & 52.6 \\ 
 & optimal& 0 &0.35 &0& 0.3 & 28.0 & 55.7  & 16.3 & 46.0 \\
  & minN& 0 &0.35 &0& 0.3  & 25.9 & 66.7 & 7.4 & 42.0 \\
\cline{1-10}
$3^\dagger$ & BOP2 & 0 & 0.35 &0 & 0.7  & 89.8 & 10.2 & 0 & 59.1 \\ 
  & optimal& 0 &0.35 &0& 0.7 & 79.2 & 10.8 & 10.0 & 47.2  \\
  &minN& 0&0.35 &0& 0.7 & 75.9 & 18.8 & 5.3 & 45.1 \\
  
\hline
\end{tabular}
 \begin{tablenotes}
	 \item[*] Treatment in experimental arm is futile
     \item[$\dagger$] Treatment in experimental arm is effective
   \end{tablenotes}
\end{center}
\end{threeparttable}
\end{table}

\begin{table}
\begin{threeparttable}
\caption{Go rate, no-go rate, consider rate and average sample size for the proposed two-arm BOP2-DC designs (optimal and minN) and the BOP2 design for single time-to-event endpoint. Experimental treatment is considered as futile when $\theta_{E}=\theta_{C}=6$ and experimental treatment is considered as effective when $\theta_{E}=10$. Maximum sample size is 90, interim analysis is conducted when sample size reaches 30, 60.}
\begin{center}
 \begin{tabular}{p{0.4cm}p{1.6cm}p{0.5cm}p{0.5cm}p{0.5cm}p{0.5cm}p{1.3cm}p{1.3cm}p{1.5cm}p{2cm}p{2cm}} 
 \hline
Sc. & Design & $\theta_{_{LRV}}$ & $\theta_{_{CMV}}$&  $\theta_{C}$&  $\theta_{E}$ & Go rate(\%) & No-go rate(\%) & Consider rate (\%) & Average sample size & Duration of trial\\
\hline
$1^*$ & BOP2& 0 & 2 &6 & 6 & 9.6 & 90.4 & 0  &64.0 & 74.3  \\ 
& optimal& 0 & 2 &6 & 6 & 5.0 & 91.2 & 3.8  &56.5 & 64.0  \\ 
  & minN&0 & 2 &6 & 6 &4.8 & 88.6 & 6.6 & 53.8 & 61.1 \\ 
\cline{1-11}
2 &BOP2& 0 & 2 &6& 7.5  & 34.3 & 65.7&0 &74.6 & 90.2\\
& optimal& 0 & 2 &6& 7.5  & 22.7 & 61.1 &16.2 & 68.9 & 81.6\\
  & minN& 0 & 2 &6& 7.5 & 22.3 & 56.0 & 21.7 & 67.1 & 79.7  \\
\cline{1-11}
$3^\dagger$ & optimal& 0 &2 &6& 10  &74.0 & 26.0 & 0 & 83.4 & 103.6 \\
& optimal& 0 &2 &6& 10  &64.5 & 18.5 & 17.0 & 76.4 &91.6\\
  & minN& 0&2 &6& 10 &63.4 & 18.4 & 18.2 & 74.9 & 89.8 \\
  
\hline
\end{tabular}
 \begin{tablenotes}
	 \item[*] Treatment in experimental arm is futile
     \item[$\dagger$] Treatment in experimental arm is effective
   \end{tablenotes}
\end{center}
\end{threeparttable}
\end{table}

\begin{table}
\begin{threeparttable}
\caption{Go rate, no-go rate, consider rate and average sample size for the proposed two-arm BOP2-DC designs (optimal and minN) and the BOP2 design for multiple endpoints. Experimental Treatment is considered as futile when $\theta_{1}=0.1$ and $\theta_{2}=0.2$, while experimental treatment is considered as effective when $\theta_{1}=0.3$ and $\theta_{2}=0.4$. Let $\theta = (\theta_{1}, \theta_{2})$. Maximum sample size is 60,  interim analysis is conducted when sample size reaches 30, 45.}
\begin{center}
 \begin{tabular}{p{0.4cm}p{1.6cm}p{1cm}p{1cm}p{1cm}p{1.4cm}p{1.5cm}p{1.5cm}p{1.5cm}p{2cm}} 
 \hline
Sc. & Design & $\theta_{_{LRV}}$ & $\theta_{_{CMV}}$&  $\theta_{C}$&  $\theta_{E}$ & Go rate (\%) & No-go rate (\%) & Consider rate (\%) & Average sample size \\
\hline
$1^*$ & BOP2 & 0,0 & 0.2,0.3&0.1,0.2  & 0.1,0.2  & 9.7 &90.3 & 0 & 36.5 \\ 
& optimal & 0,0 & 0.2,0.3&0.1,0.2  & 0.1,0.2  & 4.6 & 75.9 & 19.5 & 45.2 \\ 
  & minN & 0,0&0.2,0.3 &0.1,0.2  &0.1,0.2 & 3.2 & 92.1 & 4.7 & 35.0\\ 
\cline{1-10}
2 & BOP2& 0,0 &0.2,0.3 &0.1,0.2& 0.25,0.35  & 36.2& 63.8  & 0 & 43.3 \\
& optimal& 0,0 &0.2,0.3 &0.1,0.2& 0.25,0.35  & 63.7& 19.9  & 16.4 & 44.4 \\
  & minN& 0,0 &0.2,0.3 &0.1,0.2&0.25,0.35  & 50.2 & 39.6 & 10.2 & 45.0\\
\cline{1-10}
$3^\dagger$ & BOP2& 0,0 &0.2,0.3 &0.1,0.2& 0.3,0.4  &  46.6 & 50.4 & 0 & 46.6\\
& optimal& 0,0 &0.2,0.3 &0.1,0.2& 0.3,0.4  &  83.2 & 8.0 & 8.8 & 40.6\\
  & minN& 0,0&0.2,0.3 &0.1,0.2&0.3,0.4 & 71.0& 19.6 & 9.4 & 46.3\\
\hline
\end{tabular}
 \begin{tablenotes}
	 \item[*] Treatment in experimental arm is futile
     \item[$\dagger$] Treatment in experimental arm is effective
   \end{tablenotes}
\end{center}
\end{threeparttable}
\end{table}

\begin{table}
\begin{threeparttable}
\caption{Go rate, no-go rate, consider rate and average sample size for the proposed two-arm BOP2-DC designs (optimal and minN) and the BOP2 design for efficacy and toxicity endpoints. Experimental Treatment is considered as futile when $RR=0.2$ and $TOX=0.1$, while experimental treatment is considered as effective when $RR=0.5$ and $TOX=0.3$. let $\theta=(RR,TOX)$. Maximum sample size is 60,  interim analysis is conducted when sample size reaches 30, 45.}
\begin{center}
 \begin{tabular}{p{0.4cm}p{1.6cm}p{0.6cm}p{1.5cm}p{1.2cm}p{1cm}p{1.5cm}p{1.5cm}p{1.5cm}p{2cm}} 
 \hline
Sc. & Design & $\theta_{_{LRV}}$ & $\theta_{_{CMV}}$&  $\theta_{C}$&  $\theta_{E}$ & Go rate (\%) & No-go rate (\%) & Consider rate (\%) & Average sample size \\
\hline
$1^*$  & BOP2 & 0,0 & 0.25,0.15 &0.2,0.1 &0.2,0.1  & 10.0 & 90.0 & 0 & 39.9 \\ 
 & optimal & 0,0 & 0.25,0.15 &0.2,0.1 &0.2,0.1  &4.6 & 79.5 & 15.9 & 40.5 \\ 
  & minN & 0,0&0.25,0.15 &0.2,0.1 &0.2,0.1 & 4.6 & 94.6& 0.8 & 32.0\\ 
\cline{1-10}
2 & BOP2& 0,0 &0.25,0.15 &0.2,0.1& 0.3,0.3  & 57.9 & 42.1  & 0 & 53.0 \\
& optimal& 0,0 &0.25,0.15 &0.2,0.1& 0.3,0.3  & 44.7 & 27.7  & 27.6 & 43.3 \\
  & minN& 0,0 &0.25,0.15 &0.2,0.1&0.3,0.3  & 39.2 & 56.7 & 4.2 & 36.2\\
\cline{1-10}
$3^\dagger$ & BOP2 & 0,0 &0.25,0.15 &0.2,0.1& 0.5,0.3  &  87.0 & 13.0 & 0 & 57.9\\
 & optimal& 0,0 &0.25,0.15 &0.2,0.1& 0.5,0.3  &  88.5& 7.0 &4.5 & 34.6\\
  & minN& 0,0&0.25,0.15 &0.2,0.1&0.5,0.3 & 79.2 & 19.4 & 1.4 & 35.7\\
\hline
\end{tabular}
 \begin{tablenotes}
	 \item[*] Treatment in experimental arm is futile
     \item[$\dagger$] Treatment in experimental arm is effective
   \end{tablenotes}
\end{center}
\end{threeparttable}
\end{table}